\documentclass{jkas}

\def\year{2024} 
\def\volume{xx} 
\def\issue{xx} 
\def\beginpage{1} 
\def\received{---} 
\def\accepted{---} 
\def\published{---} 
\date{Received November 20, 2023 \received; Accepted \accepted; Published \published}

\newcommand{\msun}{M$_{\odot}$}
\newcommand{\cc}{$^{12}$C$+^{12}$C~}

\title{%
\cc Reaction Rates and the Evolution of a Massive Star
}

\author[1]{Gwangeon Seong}{0009-0004-9886-3249}
\author[2]{Yubin Kim}{0009-0009-0761-5095}
\author[1]{Kyujin Kwak}{0000-0002-2304-7798}
\author[3]{Sunghoon Ahn}{0000-0001-8190-4914}
\author[2,3]{Chaeyeon Park}{0009-0007-9796-8673}
\author[3]{Kevin Insik Hahn}{0000-0002-9250-6484}
\author[2]{Chunglee Kim}{0000-0003-3040-8456}

\affil[1]{Department of Physics, Ulsan National Institute of Science and Technology (UNIST), Ulsan 44919, Republic of Korea}
\affil[2]{Department of Physics, Ewha Womans University, Seoul 03760, Republic of Korea}
\affil[3]{Center for Exotic Nuclear Studies, Institute for Basic Science, Daejeon 34126, Republic of Korea}

\def\corrauthor{%
C. Kim, \email{chunglee.kim@ewha.ac.kr}\\
* {G.\ Seong and Y.\ Kim contributed equally to this work and share first authorship}
}

\def\runningauthor{%
Seong et al.
}

\def\runningtitle{%
\cc Reaction Rates
}

\def\keywords{%
astronomy --- astrophysics --- journals: individual: JKAS
}

\def\abstracttext{%
Carbon fusion is important to understand the late stages in the evolution of a massive star. Astronomically interesting energy ranges for the \cc reactions have been, however, poorly constrained by experiments.  
Theoretical studies on stellar evolution have relied on reaction rates that are extrapolated from those measured in higher energies. In this work, we update the carbon fusion reaction rates by fitting the astrophysical S-factor data obtained from direct measurements based on the Fowler, Caughlan, \& Zimmerman (1975) formula\nocite{Fowler1975}. We examine the evolution of a 20~\msun~star with the updated \cc reaction rates performing simulations with the MESA (Modules for Experiments for Stellar Astrophysics) code. Between 0.5 and 1 GK, the updated reaction rates are 0.35 to 0.5 times less than the rates suggested by Caughlan and Fowler (1988). The updated rates result in the increase of core temperature by about 7\% and of the neutrino cooling by about a factor of three. Moreover, the carbon-burning lifetime is reduced by a factor of 2.7. The updated carbon fusion reaction rates lead to some changes in the details of the stellar evolution model, their impact seems relatively minor compared to other uncertain physical factors like convection, overshooting, rotation, and mass-loss history. The astrophysical S-factor measurements in lower energies have large errors below the Coulomb barrier. More precise measurements in lower energies for the carbon burning would be useful to improve our study and to understand the evolution of a massive star.
}

\begin{document}
\jkashead 
\section{Introduction}
\label{sec:intro}
The reaction rate of the carbon fusion is one of the basic inputs in modeling of stellar evolution. Better precision on nuclear reactions relevant to stellar evolution would put better constraints on the properties of a core as well as the evolutionary phases of a star. Two dominant channels of the carbon fusion reaction relevant to the massive stellar evolution are given as follows
\begin{align}
^{12}\mathrm{C} + ^{12}\mathrm{C} \, \rightarrow \, ^{24}\mathrm{Mg}^{*} \, & \rightarrow \, ^{20}\mathrm{Ne} + \alpha && (Q=4.62 \, \mathrm{MeV}), \nonumber\\
& \rightarrow \, ^{23}\mathrm{Na} + p  && (Q=2.24 \, \mathrm{MeV})~. \nonumber \\
\end{align}\label{eq:channels}
The upper and lower reaction channels in Eq.\ (\ref{eq:channels}) are called $\alpha$- and $p$ (proton)-channels, because each fusion reaction produces $\alpha$ particles or protons, respectively. The branching ratio of these two reactions is almost half, i.e., $\alpha$- and $p$-channels occur with similar probabilities when two carbon nuclei fuse. In general, nuclear reaction rates depend on the center-of-the-mass energy ($E_{c.m.}$) or temperature $T$ and can be measured for each channel, separately. The \emph {total} reaction rates of the carbon fusion (ignoring other channels that have much lower branching ratios) are obtained by sums of the two rates estimated at different center-of-mass energies. In order to understand the massive star evolution, we need to determine the exact rates for \cc reactions at temperatures around $0.5-1$ GK. In other words, we are interested in the \cc reaction rates below the Gamow window that ranges from 1.5 to 2.5 MeV \citep{fowler1967, C.E.Rolfs&W.S.Rodney}. The reaction rates of the carbon fusion can be determined once the cross section $\sigma$ is known. Over the past decades, several experimental studies have been performed to measure the \cc fusion cross section directly or indirectly. In this work, we focus on direct measurements, which do not rely on theoretical models.

The main technique used in the direct measurement is the spectroscopy of charged particles and/or $\gamma$ rays. The charged particles used in direct measurements are basically $\alpha$ particles or protons emitted by the evaporated residues during the reaction. It is also possible to measure $\gamma$-rays emitted from excited states of $^{20}$Ne and $^{23}$Na in order to extract the production of $^{24}$Mg compounds via $\alpha$- or $p$-channels, respectively \citep{High1977, BarronPaloS2006, Aguilera2006}. Considering results published in literature since 1960s, $\sigma(E_{c.m.})$ obtained by direct measurements at energies larger than $\sim3$ MeV are reliable and consistent between different experiments. At lower energies below 3 MeV, however, measurement errors of $\sigma(E_{c.m.})$ for the \cc reaction are typically tens of percent or even a few hundred percent.

There are two main reasons that make direct measurement of the carbon fusion reaction extremely challenging. Firstly, the \cc cross section is expected to steeply decrease as energy decreases below the Gamow window. Secondly, the cosmic background or natural radiation increases at lower energies. 
There is a solution to overcome the second issue. \citet{Jiang2018} proposed a particle-$\gamma$ coincidence technique that could minimize the backgrounds and effectively remove ambiguities in the measurement. The main idea of this technique is to detect evaporated charged particles and $\gamma$-ray in coincidence. Although the coincidence technique can improve the significance of true events, the first issue of the low cross section remains the same. All spectroscopy-based (direct) experiments suffer significant statistical noise due to low cross section toward lower energies. 

Nuclear reaction rates are an important ingredient in the stellar evolution simulation.
The standard reaction rates of the carbon fusion used in stellar evolution simulation are given by \citet{CaughlanFowler1988} (CF88 hereafter). The rates presented by CF88 are effectively obtained from an empirical extrapolation, which is based on a generalized black-body model with a square-well potential \citep{Fowler1975}. 
Due to the lack of precise measurements available back in 1988, reaction rates below $\sim 3$ MeV of carbon fusion are poorly constrained in CF88. In spite of this caveat, the CF88 rates are in general considered as standard values in the stellar evolution simulation.

In this work, we examine the evolution of a 20 \msun~star and element production during the carbon burning with the compilation of data on the \cc rates based on the direct measurement. We extrapolated the data with the fitting formula suggested by \citet{Fowler1975} over a broad range of temperatures. The newly obtained rates are then incorporated into the stellar evolution simulation.

This work is organized as follows. In section \ref{sec:data}, we explain the data selection and how we update the \cc reaction rates. In section \ref{sec:stellar}, we describe the stellar evolution simulation and compare results based on the CF88 rates and updated rates. In section \ref{sec:discussion}, we summarize results and discuss their implications.

\begin{table}
    \centering
    \caption[]{
    A list of the direct measurement experiments used in this work. Each column represents the range in units of the center-of-mass energy, measurement method, and reference for each work. Individual datasets for each work are shown in Fig.\ \ref{fig:sfit}.}

    \begin{tabular}{ccc}
    \toprule
     $E_{c.m.}$ (MeV) & method  & Refs. \\
    \midrule
     $2.15 - 3.98$ & Particle spectroscopy & [1] \\
     $2.16 - 5.34$ & Particle-$\gamma$ coincidence & [2] \\
     $2.25 - 6.01$ & $\gamma$-ray spectroscopy &  [3] \\ 
     $2.45 - 4.91$ & Particle spectroscopy     &  [4] \\
     $2.46 - 5.88$ & $\gamma$-ray spectroscopy &  [5] \\ 
     $2.68 - 4.93$ & Particle-$\gamma$ coincidence & [6] \\
     $3.23 - 8.75$ & Particle spectroscopy     &  [7] \\ 
     $4.42 - 6.48$ & $\gamma$-ray spectroscopy &  [8] \\  

    \bottomrule
    \end{tabular}
    \label{tab:sfactor}
    \\

\small {1) \citet{Zickefoose2011}, 2) \citet{Fruet2020}, 3) \citet{BarronPaloS2006}, 4) \citet{Mazarakis1973}, 5) \citet{High1977}, 6) \citet{Jiang2018}, 7) \citet{Patterson1969}, 8) \citet{Aguilera2006}}
\end{table}

\section{Estimation of New Reaction Rates}\label{sec:data}

For a given temperature, 
the \cc reaction rate ($N_A\langle\sigma\nu\rangle$) is given as follows:  

\begin{equation}
\label{eq:reaciton} 
N_A\langle\sigma\nu\rangle={\rm A}\times 
\int_0^{\infty}\sigma(E_{c.m.})E_{c.m.}~{\rm exp}\left(-\frac{E_{c.m.}}{kT}\right) dE_{c.m.}~,
\end{equation}
where $A$ is a coefficient determined by $\left(\frac{8}{\pi\mu}\right)^{1/2}\frac{N_A}{(kT)^{3/2}}$, $N_A$ is Avogadro's number, $\mu$ is the reduced mass of the system, $k$ is the Boltzmann constant, $T$ is the temperature at which the reaction occurs, $E_{c.m.}$ is the center of mass energy. The cross section $\sigma(E_{c.m.})$ and the astrophysical S-factor $S(E_{\rm c.m})$ (S factor hereafter) are related by the equation below

\begin{equation} \label{eq:Sfactor}
    \sigma(E_{c.m.}) = S(E_{c.m.})E_{c.m.}^{-1} e^{-2\pi\eta}~,
\end{equation}
where $\eta$ is the Sommerfeld parameter. Considering the carbon burning, the power index $2\pi\eta \propto Z_1Z_2e^2/{\hbar}\nu = 87.21E_{c.m.}^{-1/2}$, where $Z_{i}$ is an atomic number, and ${\hbar}\nu$ stands for the Planck constant multiplied to the relative velocity of an incident particle in the center-of-the-mass frame. 

 In this work, we extrapolated the compilation data of the S-factor from the literature by applying the fitting formula given in \citet{Fowler1975}. The updated $S(E_{c.m.})$ is then converted to $\sigma(E_{c.m.})$ so that we could compute the reaction rates using Eq.\ (\ref{eq:reaciton}).
\begin{figure*}[t]
    \centering
    \includegraphics[width=\textwidth]{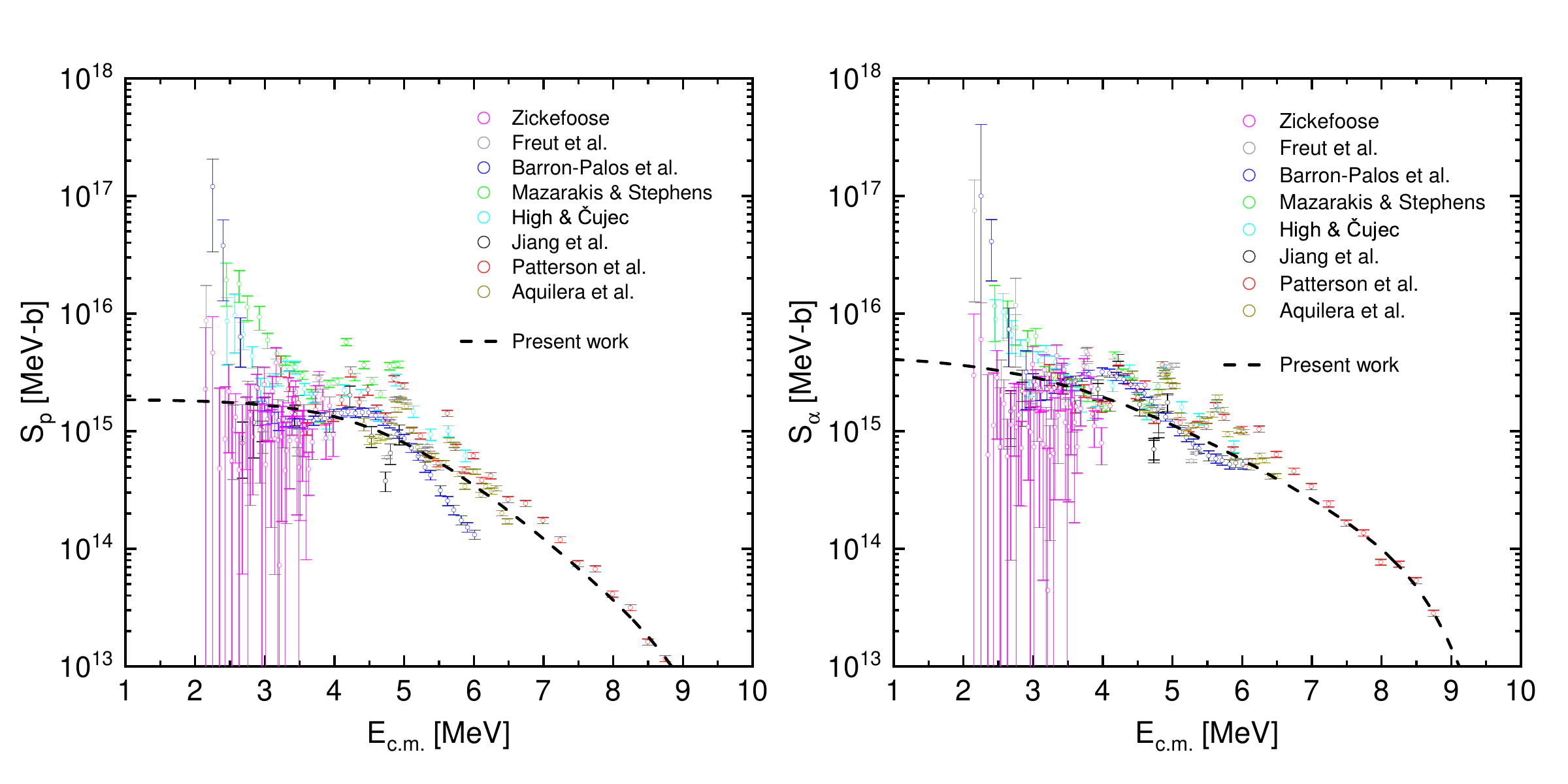}
    \caption{Fitted $S(E_{c.m.})$ distributions obtained from this work are shown as black dashed lines. The fitted distribution are overlaid with the  direct measurement data from literature (listed in Table 1) for the $p$-channel (left, $S_{p}$) and the $\alpha$-channel (right, $S_{\alpha}$), respectively. 
    }
    \label{fig:sfit}
\end{figure*}
In order for performing the extrapolation, we have collected the S-factor data directly measured in the energy range of $2.15-8.745$ MeV. If the data are given as $\sigma(E_{c.m.})$, we converted them to the corresponding S-factor values using Eq.\ (\ref{eq:Sfactor}). 

Table \ref{tab:sfactor} presents energy ranges, methods of measurements, and references for the eight experiments used in this work. We adopted experimental data of the cross section $\sigma(E_{c.m.})$ 
from literature only when there are tabulated data for $\alpha$- and $p$-channels, respectively. 
Most work in Table \ref{tab:sfactor} provides the mean and percentage errors for each data. If only a range of measured values was given \citep[e.g.\ low-energy data of][]{Fruet2020}, we calculated an arithmetic mean between the lower and upper limits given in the range assuming a symmetric error.

\subsection{Extrapolation of S-factor}
\label{subsec:extrapolation}
Despite many experimental efforts to measure the cross section of \cc reaction directly, reliable data in the low energy range are not only insufficient but also have poor precision. So it is crucial to extrapolate the S-factor down to the Gamow window. The fitting formula suggested by \citet{Fowler1975} is written as follow
\begin{equation} \label{eq:fitting_eq} 
S(E)= \frac{S(0){\rm exp}(-\alpha E)}{{\rm exp}(-\gamma E^m)+ b~{\rm exp}(+\beta E)}~.
\end{equation}
This equation contains six parameters and approaches $S(0)\rm exp(-\alpha E)$ at low energy and to $(1/b) S(0)\rm exp[-(\alpha+\beta)E]$ at high energy, respectively. The transition behavior of the negative slope around 5 MeV is determined by the term $\rm exp(-\gamma E^m)$. The cross section and the corresponding S-factor expressed in Eq.\ (4) are the outcomes of a model where two charged nuclei (two carbon nuclei in this work) react (or fuse) to form a merged nucleus under the two combined potentials, repulsive Coulomb potential at large separation and attractive square-well potential at small separation which approximates the attractive strong (nuclear) force.

One of the main objectives of this work is to present the differences in $S(E_{c.m.})$ distributions based on the original CF88 results versus that is based on data obtained from direct measurements to date. Figure \ref{fig:sfit} shows the best-fit results for $S(E_{c.m.})$ for $\alpha$- and $p$-channels, respectively. Data from eight references listed in Table 1 are presented in different colors. Let us also note that gamma-ray spectroscopy technique (experiments) have more data points than the coincidence technique at low energies. This can be easily confirmed by Figure 1. Generally speaking, the more data points there are, the better the fitting quality. This is the main reason we consider all spectroscopy data as listed in Table \ref{tab:rates}.

As expected, the S-factor at high energy is not significantly different from the original curve of CF88. As pointed out in many previous works, $S(E_{c.m.})$ values toward lower energies are poorly constrained due to large uncertainties in measurements. As expected, our best-fit curve is similar to CF88. Let us point out that the fitted values of the S-factor are slightly \emph{smaller} than those of CF88 in the astrophysically interesting temperature range. Based on the results that are described in later chapters, we confirmed that some of the physical properties of stellar evolution are sensitive to the \cc rate.

\begin{figure}[t]
    \centering
    \includegraphics[width=8.5cm]{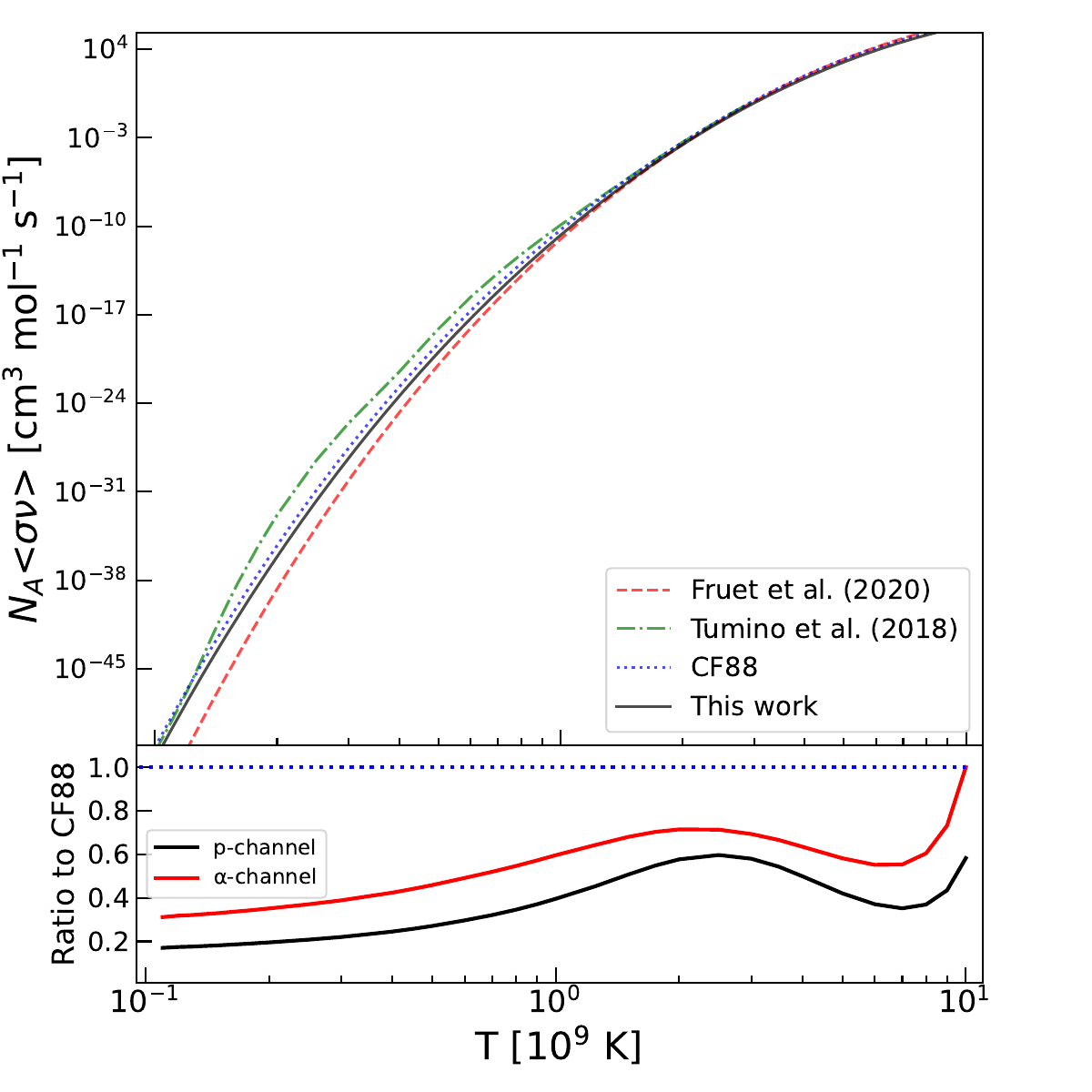}
    \caption{Comparisons of the $\mathrm{^{12}C+^{12}C}$ reaction rates obtained from this work and previous studies. The top panel displays the total reaction rates obtained from our fitting results (black solid line); CF88 (blue dotted line), the hindrance model in \citet{Fruet2020} (red dashed line), and \citet{Tumino2018} (green dash-dotted line). The lower panel shows ratios of the reaction rates of this work to those in CF88. The black solid line and red solid line correspond to the ratios obtained for $\alpha$- and $p$-channels, respectively.}
    \label{fig:rates}
\end{figure}

In addition to the fitting results, we also present a Table of the reaction rates of the carbon fusion between $T=[0.1, 5]$ GK in Appendix A.  We calculated proton- and $\alpha$-channels separately and then added them for calculating the total rates at different temperature. The astronomically interesting range of temperatures is chosen in the context of the evolution of a massive star. Figure \ref{fig:rates} compares the reaction rates obtained from this work and previous works (top panel). As a comparison, we also show the ratio of our rate to the rate of CF88 for $\alpha$- and $p$- channels, respectively (bottom panel). The reaction rates obtained from this work are lower than those suggested by CF88 slightly. The difference is more apparent between our work and the result in \citet{Tumino2018} that prefers tens of up to thousands larger values of the carbon-burning reaction rates. We note that \citet{Mukhamedzhanov2019} obtained more consistent results with CF88 when applying different renormalization for the data obtained by \citet{Tumino2018}. The indirect measurements show larger baseline and resonance features and it is understandable that \citet{Tumino2018}'s rates are larger than any work based on direct measurements. The $\mathrm{\alpha}$-channel rates obtained from this work (red curve in the bottom panel) are approximately 2 times larger than the $p$-channel rates at $0.5 - 1$ GK

\section{Stellar Evolution}\label{sec:stellar}
The carbon burning is expected to occur during the evolution of massive stars with a mass of approximately 8 times or more than that of the Sun. A typical approach to study a massive star evolution is to utilize computer simulation. In this study, we employ Modules for Experiment for Stellar Astrophysics ({\tt MESA} hereafter, revision r12115) \citep{Paxton2011, Paxton2013, Paxton2015, Paxton2018, Paxton2019, Jermyn2023}, which is an open source 1D stellar evolution code. We simulated the evolution of a non-rotating, mass-losing star with an initial mass of 20 \msun~until the carbon abundance in the core reached 0.001 at the completion of the carbon-burning phase. We examined the influence of the new carbon-burning reaction rates on a model star. We compared two models: (a) our model based on the updated $\mathrm{^{12}C+^{12}C}$ reaction rates and (b) the CF88 model refers to the stellar evolution utilizing the reaction rates given by CF88.

Model parameters are based on the solar calibration results of \citet{Farag2020}. For example, we set a mixing length $\mathrm{\alpha = 2.12}$ and an exponential overshooting \citep{Herwig2000} parameter $f_{ov} = 0.016$ representing parameterized convection. We also set the initial helium abundance (Y) to be 0.265 and the initial metallicity $Z$ to be 0.0151, taking into account the solar metallicity from \citet{Asplund2009} ($Z/X=0.0181$). To account for mass loss, we use the {\tt Dutch} scheme with a scaling factor of 1.0, which combines the mass loss schemes from \citet{Vink2001} for hot stars and \citet{deJager1988} for cool stars. Nuclear reactions of a star in the simulation are computed by the large reaction network {\tt mesa\_204} with nuclear reaction rates from the Joint Institute for Nuclear Astrophysics (JINA) REACLIB database. The current standard version is dated as 
of October 20, 2017 \citep{Cyburt2010}, except the carbon-burning reaction rates. We note two models (our model versus CF88 model) are identical except for the carbon-burning rates. 

\begin{figure}[t]
    \centering
    \includegraphics[width=8.5cm]{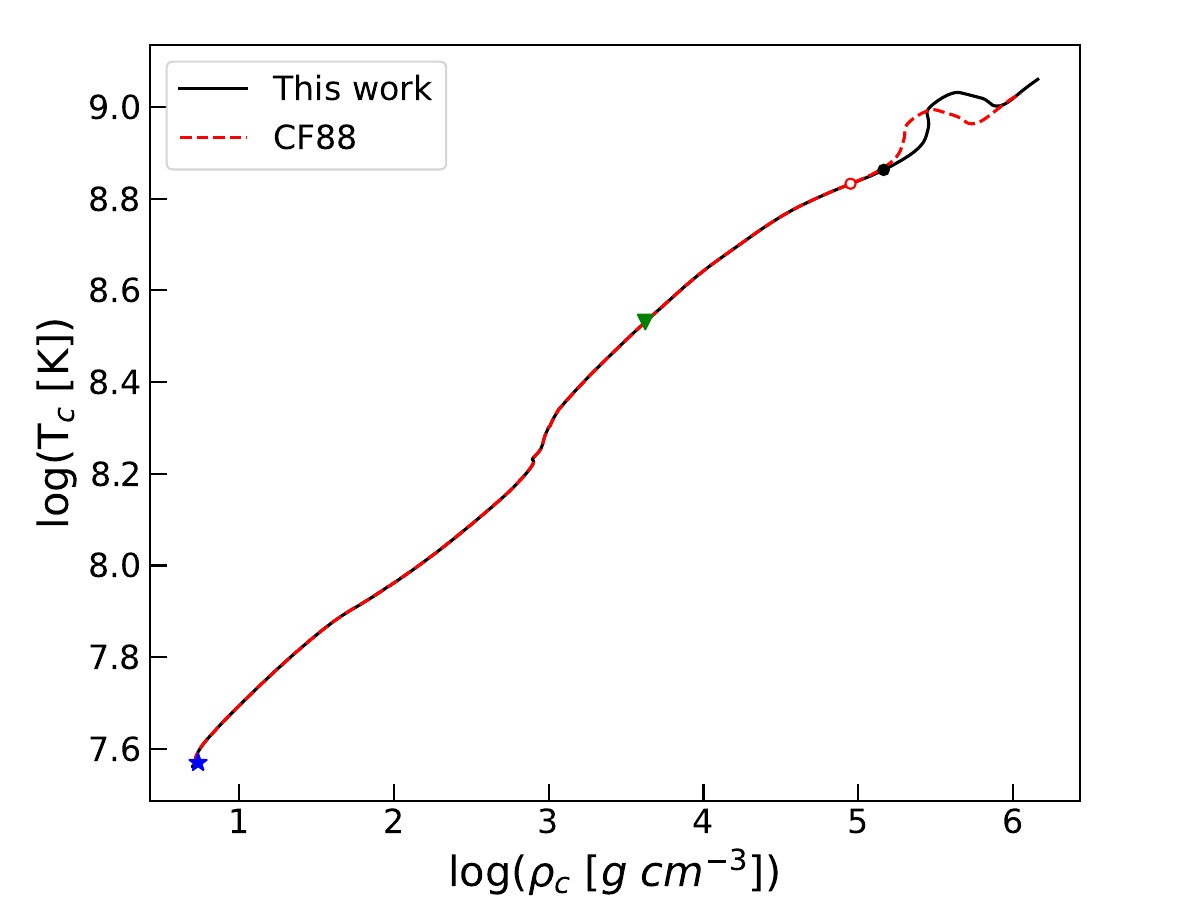}
    \caption{Evolutionary tracks in a $T_{\rm c}-\rho_{\rm c}$ plane for a 20 \msun~obtained with two reaction rates: this work (black solid line) and CF88 (red dashed line). The evolution starts from the main sequence (blue star) and continues until the end of the carbon-burning phase. The green triangle marks the point when the core He is depleted and the black and red circles represent the onset of carbon-burning for each model.}
    \label{fig:trho}
\end{figure}

Figure \ref{fig:trho} shows the evolution of core temperature and density of the 20 \msun~star obtained from our model and the CF88 model. Initially, the two models follow a similar path from the main sequence (indicated by a blue star) until the helium (He) in the core is depleted (the green triangle in the figure). As the core contracts without nuclear reactions, the core temperature and density increase until the carbon-burning phase begins. We define the beginning of the carbon-burning phase as the point where 1\% of carbon in the core has burned. As shown in the figure, our reaction rates require higher temperatures and densities for the carbon burning than what was expected in the CF88 model.

\begin{figure}[t]
    \centering
    \includegraphics[width=8.5cm]{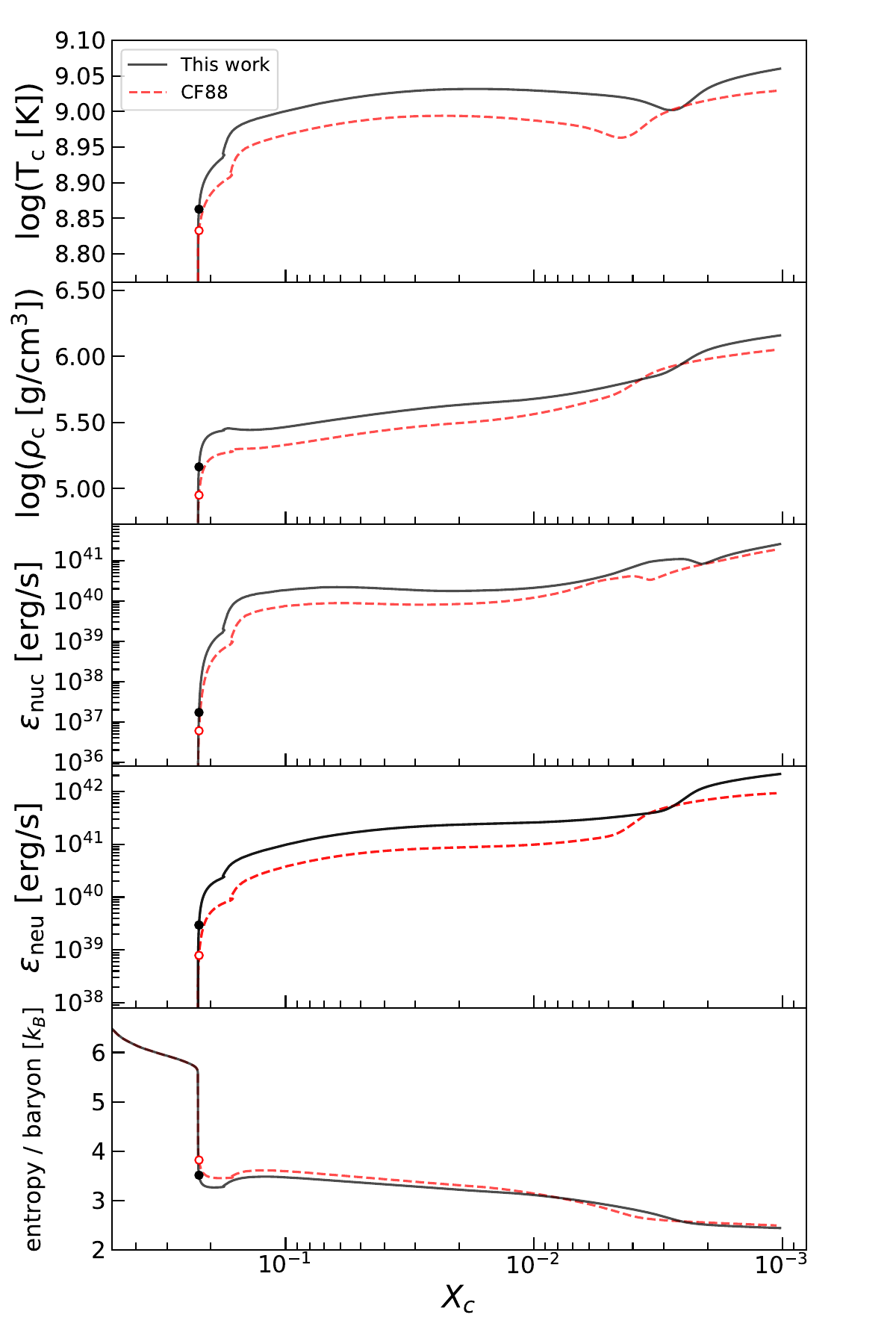}
    \caption{Evolution of core properties as a function of $X_{c}$. Selected physical quantities are shown: the core temperature~$T_{\rm c}$ (top panel), core density~$\mathrm{\rho_{c}}$ (second panel), total power from the carbon burning~$\mathrm{\varepsilon_{nuc}}$ (third panel), total power emitted in neutrinos~$\mathrm{\varepsilon_{neu}}$ (fourth panel), and central entropy per baryon (bottom panel). In all panels, this work and CF88 rates are presented with black solid and red dashed lines, respectively. The black and red circles in all panels represent each physical quantity at the beginning of the carbon-burning phase.} 
    \label{fig:properties}
\end{figure}

Detailed properties of our model and CF88 model during the carbon-burning phase are shown in Figure \ref{fig:properties}. Note that $X_c$ is the central mass fraction of $\mathrm{^{12}C}$, i.e., the ratio of $\mathrm{^{12}C}$ mass to total mass at the stellar center. In comparison to the CF88 model, our model exhibits an approximate 7\% increase in core temperature (top panel). Core density increases 65\% at the onset of the carbon-burning phase and 28\% increase at the end of the carbon-burning phase (2nd panel). The enhanced core temperature and density result in more energy generation in the same burning phase although our reaction rates may be lower than CF88's at the same temperature (3rd panel).

Neutrino loss plays an important role in the energy budget of a star during the carbon-burning phase \citep{Woosley2002}. In the carbon-burning phase, neutrinos are mostly produced by thermal processes, dominantly electron-positron pair annihilation. The rate of neutrino production from these thermal processes mainly depends on the core temperature and electron density ($\rho_{\rm c} Y_{\rm e}$) \citep{Itoh1996}. Therefore, the amount of neutrino production depends on the changes in core temperature and density caused by nuclear reaction rates. Our model has a relatively higher temperature and density than the CF88 model. As a result, neutrino loss increases by approximately a factor of three in comparison with the CF88 model (see the bottom panel of Figure \ref{fig:properties}).

In the context of massive star evolution, central entropy of a star is tend to decrease toward later stages. This is primarily induced by complex interplay between neutrino emission, energy generation by nuclear reactions, and in particular by convection \citep{Woosley2002}. We presented the evolution of the central entropy per baryon as a function of $X_c$ in Figure 4. The expected decrease in the central entropy is confirmed. However, the evolution of the central entropy is not sensitive by carbon burning rates comparing to other properties (shown in the other panels in Fig 4).

\begin{figure}[t]
    \centering
    \includegraphics[width=8.5cm]{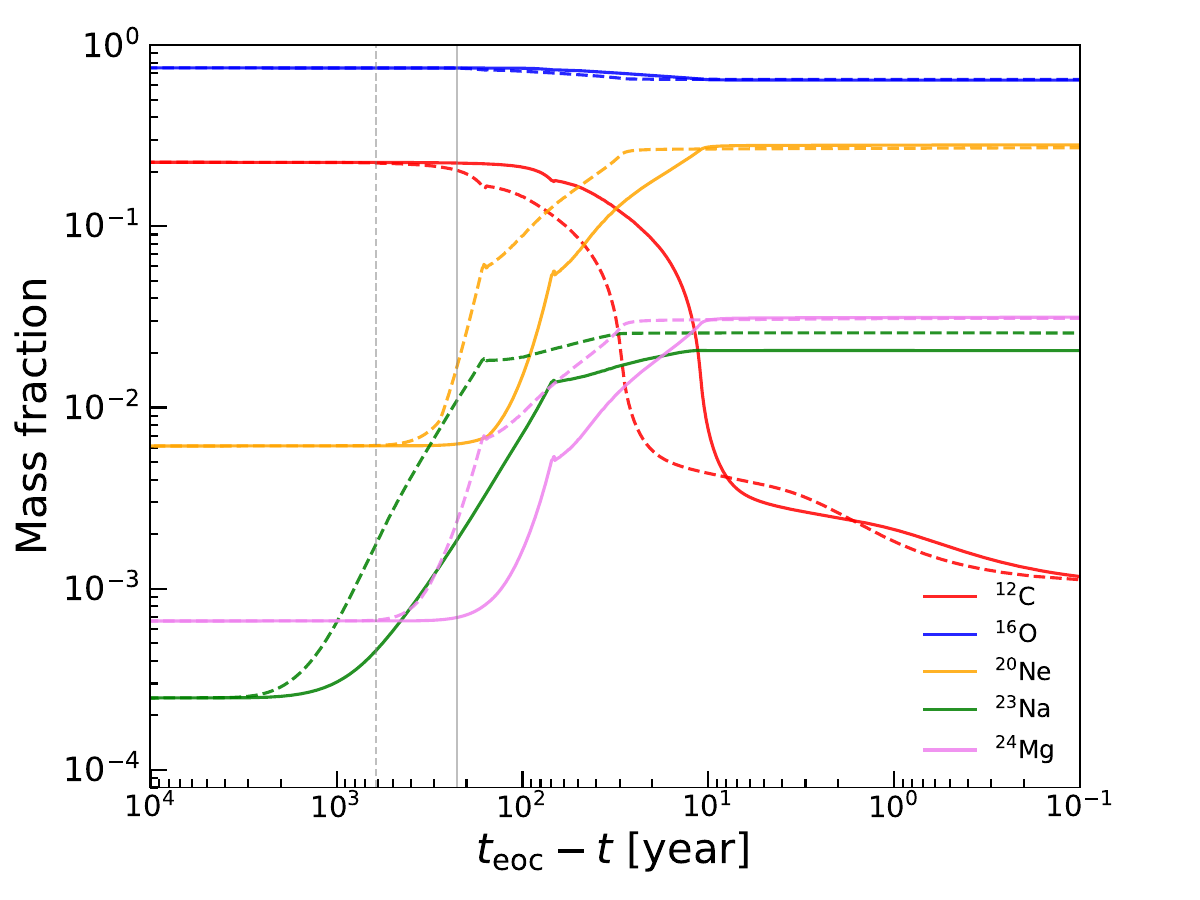}
    \caption{Evolution of chemical abundances at the center of a 20 \msun~star until the end of carbon-burning phase. Note that $t_{eoc}$ is the time to the end of carbon burning phase ($X_{c} \leq 10^{-3}$).} Results obtained from this work (solid lines) and those from the CF88 model (dashed lines) are overlaid for comparison. Gray vertical lines indicate the times at the beginning of the carbon-burning phase for each model.
    \label{fig:abundance}
\end{figure}

Figure \ref{fig:abundance} compares the abundances by mass at the center obtained from our model and from the CF88 model during the carbon-burning phase. In addition to the most abundant elements such as $^{12}$C and $^{16}$O, we show the mass fractions of $^{20}$Ne, $^{23}$Na, $^{24}$Mg. These are produced by the \cc reactions. 
Initial abundances of all other elements except $^{23}$Na are almost the same. This is expected as the two models use the same nuclear reaction rates except the carbon burning (see the bottom panel of figure \ref{fig:rates}). However, the mass fractions of $^{23}$Na are different. This can be understood by the differences in the branching ratios based on the \cc rates (see Table \ref{tab:rates}). Specifically, our updated reaction rates have a larger branching ratio for the $\alpha$-channel than that of $p$-channel. The branching ratios of the CF88 rates for the $p$- and $\alpha$-channels are 44.6\% and 55.4\%, respectively.

Our model (solid lines) predicts a longer period of core contraction due to relatively lower nuclear reaction rates compared to those from the CF88 model. Furthermore, because of the relatively higher core temperature and density, the carbon in the core is rapidly burned over a shorter period in our model. As a result, the lifetime of the carbon-burning phase with our updated reaction rates decreases by a factor of 2.7 in comparison with the CF88 model.

\begin{figure*}[t]
\centering
\begin{tabular}{cc}
    \includegraphics[width=8.5cm]{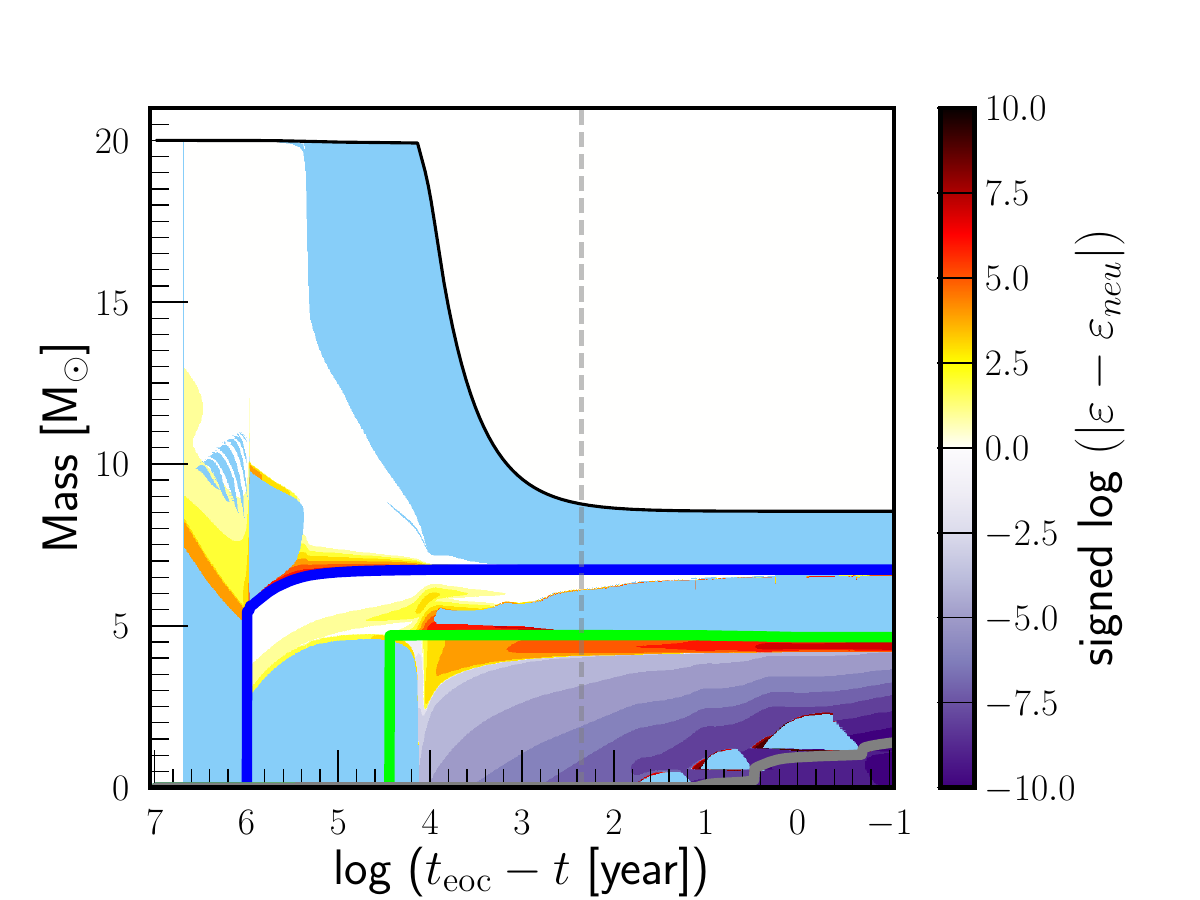}&
    \includegraphics[width=8.5cm]{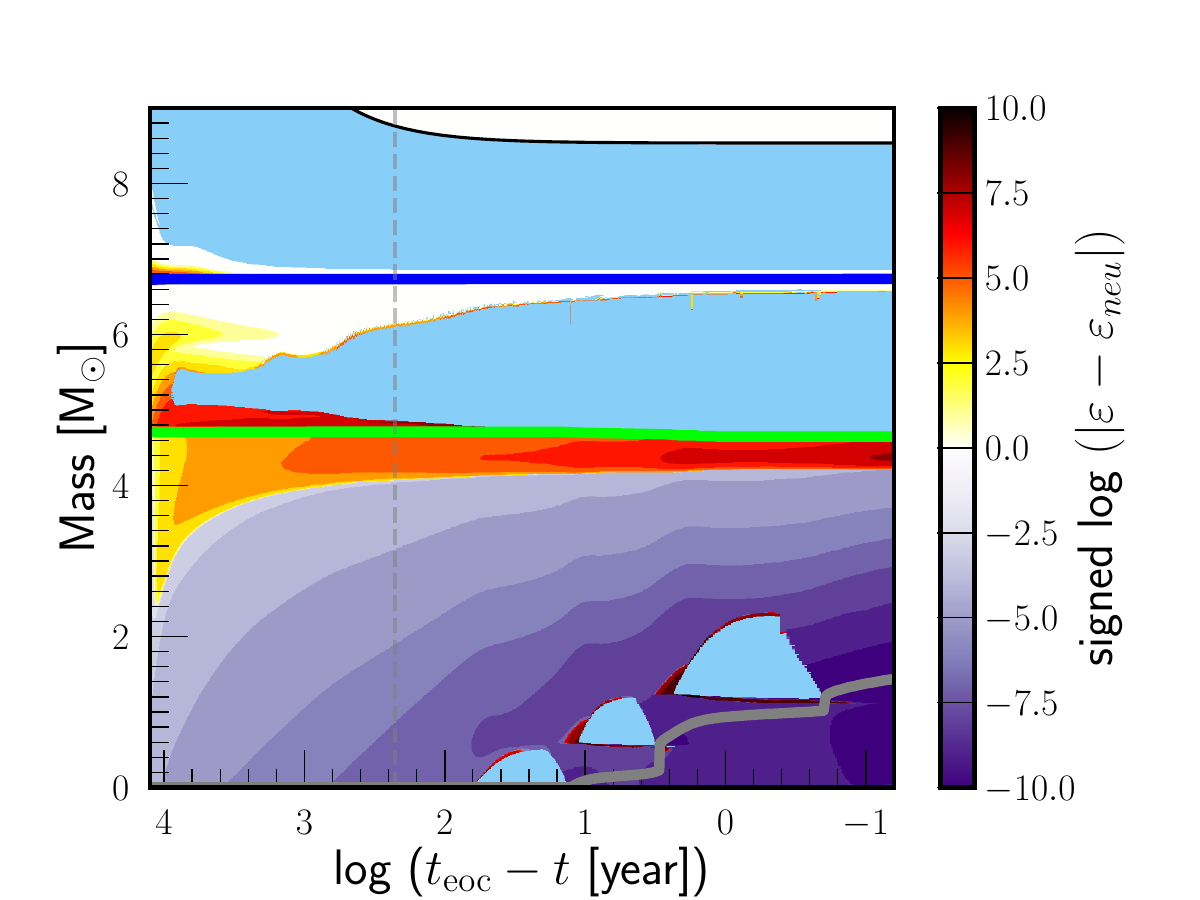}\\
    \includegraphics[width=8.5cm]{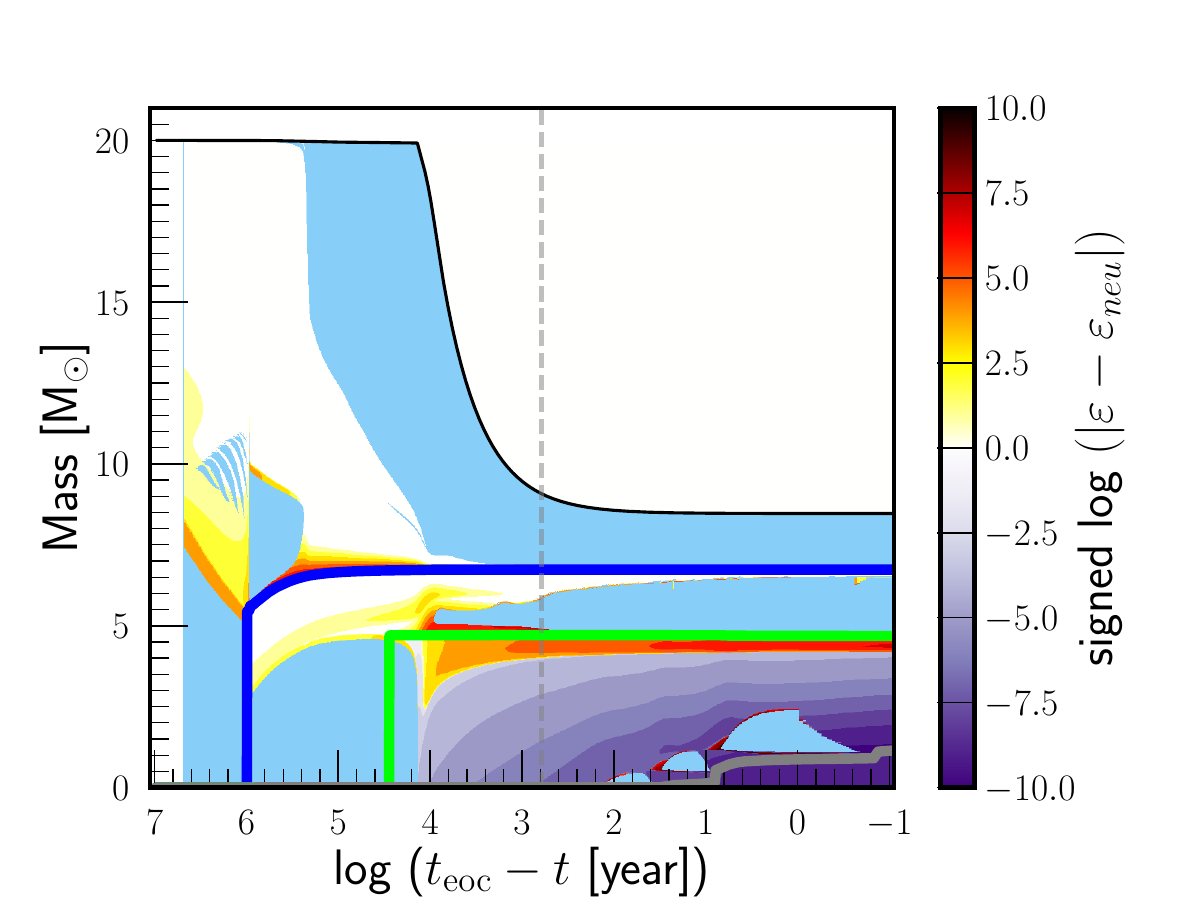}&
    \includegraphics[width=8.5cm]{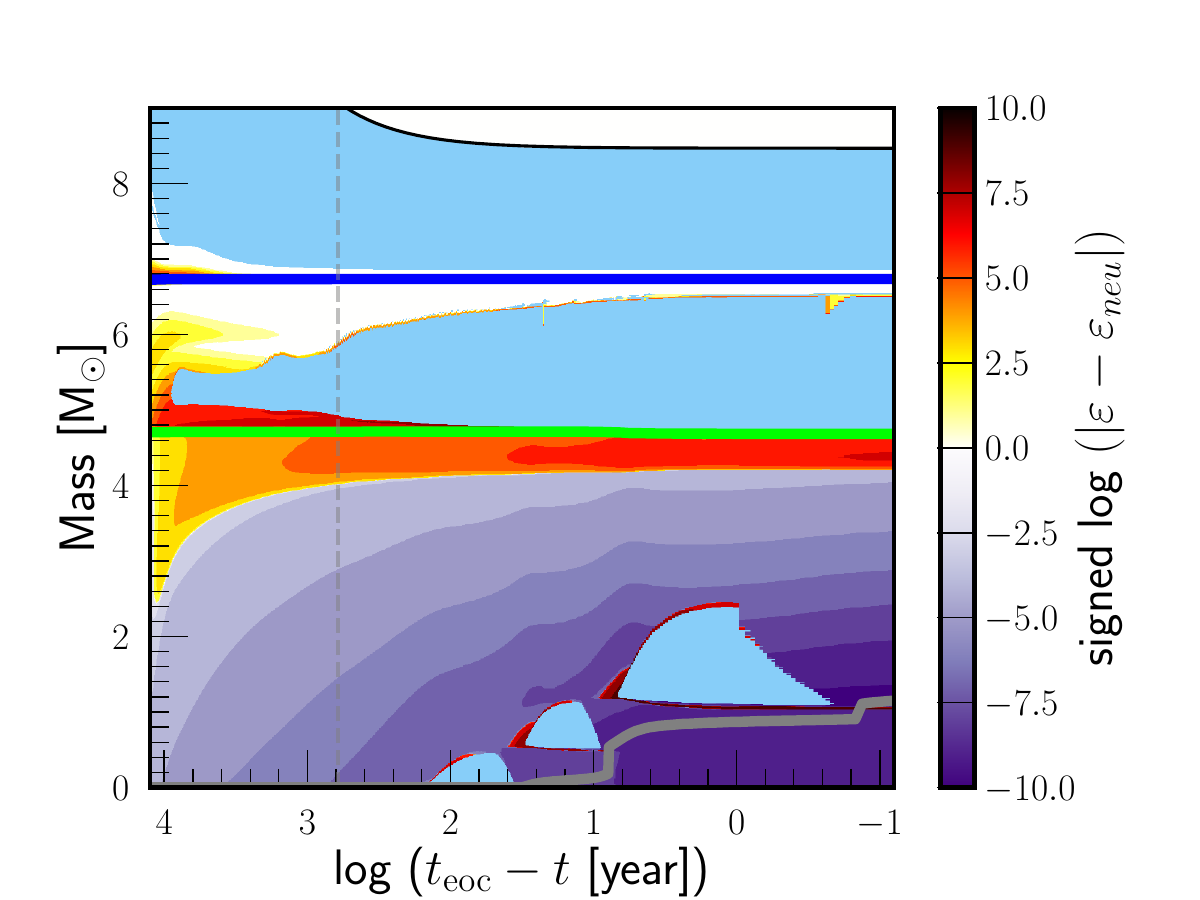}
\end{tabular}
    \caption{Kippenhahn diagrams for a 20 \msun star based on different \cc rates based on this work (top panels) and CF88 (bottom panels). Results for two time scales are presented, i.e. the entire evolution time (left column) and the last $10^{4}$ years before the end of the carbon burning phase (right column), respectively. The gray vertical dashed lines indicate the time at the beginning of the carbon burning phase. The boundaries of the He (blue), C/O (green), and O/Ne (gray) core/shell are indicated by solid lines. The total energy ($\varepsilon - \varepsilon_{neu}$), i.e., the sum of energy generation by nuclear reactions and energy loss by neutrinos, is denoted by the color bar. Convection zones are presented in sky blue.}
    \label{fig:kipp}
\end{figure*}

The development of convective cores or shells during the carbon-burning phase affects the final fate of a massive star \citep{Cheiffi2020}. The interplay between energy production from nuclear reactions and energy loss by neutrino emission plays a crucial role in the development of the convective zone. Previous works on stellar evolution simulations found that \cc reaction rates significantly impact the size of the convective carbon core and the number of convective carbon shells. This impact has been reported for models with different initial masses but it is in particular apparent for stellar models around 20 \msun~(see \citet{Woosley2002} and references therein). For example, \citet{Pignatari2013} demonstrated that  lower carbon burning reaction rates lead to a tiny or no convective core, whereas higher rates result in an expanded one in the evolution of a 25 \msun~ star.

Figure \ref{fig:kipp} compares Kippenhahn diagrams obtained from this work with those based on the CF88 rates. It prepresents the convective zones (shown in sky blue) and the areas of energy generation or emission relative to each model's mass versus evolutionary time. Note that negative values of $signed~ log (|\varepsilon - \varepsilon_{neu}$|) imply that the energy emission from neutrinos exceeds the energy produced by nuclear reactions. We also note that our results do not show a reduction in the convective carbon core/shell that is expected from the CF88 model. 
Among those we examined in this work, central entropy and convection zone are least affected by the changes in \cc rates.

\section{Summary and Discussion}
\label{sec:discussion}

In this work, we re-evaluated the carbon-burning reaction rate by carefully selecting the tabulated data of the astrophysical S-factor from the literature. We fitted the data for the $\alpha$- and $p$-channels and calculated the reaction rates for the two channels. The \cc reaction rates from our fitting results have values ranging from 0.35 to 0.5 times those of CF88 between 0.5 and 1 GK. We compare our results with previous studies. 

Applying the updated reaction rates, we performed a stellar evolution simulation for a 20 \msun~model until the end of the carbon-burning phase. We found the lifetime of the carbon-burning phase with updated rate estimates is decreased by a factor of 2.7 compared to that obtained from the CF88 model. This decrease is attributed to the higher temperature and density in the core due to the lower reaction rates. This also implies increased neutrino loss during the carbon-burning phase.

The shape of $S(E_{c.m.})$ below the Gamow window is important in astronomy but is inconclusive \citep[e.g.,][]{Becker1981, Zickefoose2011, Jiang2018}. For example, some works prefer significant increase of $S(E_{c.m.})$ below the Coulomb barrier \citep{Mazarakis1973, High1977, BarronPaloS2006}. Measurements utilizing $\gamma$-ray detection such as \citet{BarronPaloS2006} and \citet{Spillane2007} obtained more than one order of magnitude larger S-factor estimates at lower energies than those measured in higher energies. They interpreted the increase of the reaction rate toward the lower energies as evidence for the existence of a resonance near 2 MeV. 
\citet{Jiang2007} suggested the measurements broadly support the hindrance model and their results do not support the existence of resonance near 2 MeV. Recent work based on particle-$\gamma$ coincidence \citep{Jiang2018, Fruet2020, Tan2020} also reported a declining tendency in $S(E_{c.m.})$ below a few MeV. \citet{Tan2020} reported they have negative results for the existence of resonance as reported by \citet{Tumino2018}. We note the behavior of $S(E_{c.m.})$ toward low energies based on direct experiments are poorly constrained, i.e., resonance or hindrance signatures are inconclusive. In order to constrain the $S(E_{c.m.})$ toward lower energies, it is important to increase the number of true events and to reduce the statistical uncertainties in results.

We could not include some of the previous works. For example, \citet{Becker1981}, \citet{Spillane2007}, and \citet{Tan2020} are not included in this work as they do not provide tabulated data. We note the qualitative tendencies and uncertainties of the \emph{baseline} of the S-factor estimates in lower energies are consistent between the excluded works and those considered in this work. Therefore, we expect that including the excluded data does not change our main results significantly.
\citet{Tumino2018} reported a clear signature of resonances below the Coulomb barrier using a new technique, but This work is also excluded because their measurements are indirect. Although very interesting and promising for studying the reaction rates in the astrophysically interesting range, their data show large uncertainties and it is beyond the scope of this work to compare the data quality between direct and indirect measurements in order to calculate reaction rates.

In order to overcome the experimental limitations due to the low counting rate, an indirect measurement for \cc reaction was performed using the recently developed Trojan Horse Method (THM). The THM measures the three-body breakup reaction occurring at an energy above the Coulomb barrier to understand the two-body reaction of interest. Then, using a model-dependent kinematic constraint, cross sections of the two-body reaction can be extracted as a result of the theoretical description in certain approximations. Thus, in principle, the THM can study cross sections of a nuclear reaction below the Coulomb barrier all the way down to E$_{c.m.}\sim0$. 
\citet{Tumino2018} studied resonances of the \cc cross section via the THM in the energy region near the Gamow window for massive stars. Although the method is promising, however, the \cc reaction rate estimates based on THM show differences from the direct measurements up to two orders of magnitude. Therefore, we excluded their work.

We do not consider the resonances in our fitting because of two reasons. Firstly, the large uncertainties in the S-factor obtained from direct measurements make it difficult to test the effects of resonances. Secondly, we mostly focus on the reaction rate contributed by continuum levels at around 1 MeV. In other words, we are interested in the general shape of the $S(E_{\rm c.m.})$ curve. Moreover, we attempted to model-independent tendency of the carbon reaction rates. We note more recent studies on the late stage of a massive star or pre-supernovae properties have tried different rates expected by the Hindrance model with or without resonances. For example, \citet{Gasques2007} studied the consequences of strongly reduced \cc reaction rates from the work by \citet{Jiang2007} on the evolution of 20~\msun~and 60~\msun~stars. \citet{Bennett2012} discussed stellar models with various initial masses with higher reaction rates. Based on these works, \citet{Pignatari2013} presented models for a 25 \msun~star. They consider a range of \cc rate uncertainties from known measurements, which ranges from at least 20 times smaller to  $\sim50,000$ times larger than that of CF88 at $5\times10^{8}$ K. The authors considered how the rate uncertainties of \cc reaction rates affect on s- and p-processes in the evolution of massive stars. Recently, \citet{Monpribat2022} examined the influence of new reaction rates on 12 \msun~and 25 \msun~models using the {\tt GENEC} stellar evolution code \citep{Eggenberger2008}. They extrapolated the data from \citet{Fruet2020} and compared results for three model assumptions, i.e. reaction rates following the CF88 rates and rates from a hindrance model with and without resonances. Details of the hindrance models and discussion are described in \citet{Monpribat2022}.

Studies on carbon burning can provide rich information on stellar evolution, thermonuclear-powered phenomena, and neutrino emission relevant to massive stars. Any changes in the carbon-burning rates affect nucleosyntheses in the later evolution of a star, in particular a pre-supernova stage, including oxygen-burning, neon-burning, and silicon-burning phases. The updated carbon fusion reaction rates lead to some changes in the details of the stellar evolution model, their impact seems relatively minor compared to other uncertain physical factors like convection, overshooting, rotation,
and mass-loss history.

Sensitivity study for astronomical phenomenon is only possible when the data are well constrained. More measurements with higher energy resolution below the Coulomb barrier are important to pin down the shape of $S(E_{c.m.})$, and hence, the \cc reaction rates. Future experiments with better precision and/or accuracies can be very helpful to stringent constraints on the \cc reaction rates and to understand the fate of a massive star.


\acknowledgments

This work is supported by KASI 2023181002 ``Development of Cutting-Edge Technology of Laser Interferometer for International Collaboration for Next-Generation Gravitational Detectors.'' The authors also acknowledge the support from the Institute for Basic Science (IBS) of the Republic of Korea (Grants No.\ IBS-R031-D1) and the National Research Foundation of 
Korea (NRF) grant funded by the Korean government (MIST) 
(Grants No.\ 2022R1F1A1070060).

\bibliography{references.bib}

\appendix
\section{\cc reaction rates table} \label{sec:appendA}

\begin{table}[]
\centering
\begin{footnotesize}
\caption[]{\small{Reaction rates ($\mathrm{cm^{3}\ s^{-1}\ mol^{-1}}$) and branching ratios for the $p$-channel (2nd col.) and $\alpha$-channel (3rd col.), computed at different temperatures $T$ based on Eq.\ (\ref{eq:reaciton}) and Fig.\ \ref{fig:sfit}.}}
\begin{tabular}{cccc}
\toprule
$T$   {[}GK{]} & $p$-channel & $\mathrm{\alpha}$-channel & $p$ : $\alpha$ [\%] \\
\midrule
0.10  & 2.06E-53 & 4.71E-53 & 30.4 : 69.6 \\
0.11 & 5.64E-51 & 1.28E-50 & 30.6 : 69.4\\
0.12 & 8.08E-49 & 1.83E-48 & 30.6 : 69.4\\
0.13 & 6.85E-47 & 1.55E-46 & 30.6 : 69.4\\
0.14 & 3.76E-45 & 8.50E-45 & 30.7 : 69.3\\
0.15 & 1.43E-43 & 3.23E-43 & 30.7 : 69.3\\
0.16 & 4.00E-42 & 9.00E-42 & 30.8 : 69.2\\
0.17 & 8.55E-41 & 1.92E-40 & 30.8 : 69.2\\
0.18 & 1.44E-39 & 3.25E-39 & 30.7 : 69.3\\
0.19 & 2.00E-38 & 4.48E-38 & 30.9 : 69.1\\
0.2 & 2.31E-37 & 5.18E-37 & 30.8 : 69.2\\
0.21 & 2.28E-36 & 5.10E-36 & 30.9 : 69.1\\
0.22 & 1.96E-35 & 4.36E-35 & 31.0 : 69.0\\
0.23 & 1.48E-34 & 3.29E-34 & 31.0 : 69.0\\
0.24 & 9.96E-34 & 2.21E-33 & 31.1 : 68.9\\
0.25 & 6.04E-33 & 1.34E-32 & 31.1 : 68.9\\
0.26 & 3.34E-32 & 7.38E-32 & 31.2 : 68.8\\
0.27 & 1.69E-31 & 3.73E-31 & 31.2 : 68.8\\
0.28 & 7.94E-31 & 1.75E-30 & 31.2 : 68.8\\
0.29 & 3.46E-30 & 7.61E-30 & 31.3 : 68.7\\
0.3 & 1.41E-29 & 3.10E-29 & 31.3 : 68.7\\
0.35 & 6.93E-27 & 1.50E-26 & 31.6 : 68.4\\
0.4 & 1.14E-24 & 2.47E-24 & 31.6 : 68.4\\
0.45 & 8.61E-23 & 1.83E-22 & 32.0 : 68.0\\
0.5 & 3.55E-21 & 7.47E-21 & 32.2 : 67.8\\
0.55 & 9.16E-20 & 1.91E-19 & 32.4 : 67.6\\
0.6 & 1.62E-18 & 3.35E-18 & 32.6 : 67.4\\
0.65 & 2.12E-17 & 4.33E-17 & 32.9 : 67.1\\
0.7 & 2.16E-16 & 4.35E-16 & 33.2 : 66.8\\
0.75 & 1.77E-15 & 3.53E-15 & 33.4 : 66.6\\
0.8 & 1.21E-14 & 2.39E-14 & 33.6 : 66.4\\
0.85 & 7.12E-14 & 1.38E-13 & 34.0 : 66.0\\
0.9 & 3.65E-13 & 6.99E-13 & 34.3 : 65.7\\
0.95 & 1.66E-12 & 3.14E-12 & 34.6 : 65.4\\
1 & 6.80E-12 & 1.27E-11 & 34.9 : 65.1\\
1.05 & 2.54E-11 & 4.70E-11 & 35.1 : 64.9\\
1.1 & 8.75E-11 & 1.59E-10 & 35.5 : 64.5\\
1.15 & 2.79E-10 & 5.03E-10 & 35.7 : 64.3\\
1.2 & 8.34E-10 & 1.48E-09 & 36.0 : 64.0\\
1.25 & 2.34E-09 & 4.13E-09 & 36.2 : 63.8\\
1.3 & 6.23E-09 & 1.08E-08 & 36.6 : 63.4\\
1.35 & 1.57E-08 & 2.71E-08 & 36.7 : 63.3\\
1.4 & 3.80E-08 & 6.47E-08 & 37.0 : 63.0\\
1.45 & 8.79E-08 & 1.48E-07 & 37.3 : 62.7\\
1.5 & 1.95E-07 & 3.26E-07 & 37.4 : 62.6\\
1.75 & 6.45E-06 & 1.03E-05 & 38.5 : 61.5\\
2 & 1.11E-04 & 1.71E-04 & 39.4 : 60.6\\
2.5 & 8.88E-03 & 1.32E-02 & 40.2 : 59.8\\
3 & 2.23E-01 & 3.5E-01 & 38.9: 61.1\\
3.5 & 2.78E+00 & 4.05E+00 & 38.9 : 61.1\\
4 & 1.97E+01 & 3.01E+01 & 39.6 : 60.4\\
5 & 3.58E+02 & 6.15E+02 & 36.8 : 63.2\\
\bottomrule
\end{tabular}\label{tab:rates}
\end{footnotesize}
\end{table}

\end{document}